\documentclass[a4paper,11pt]{article}
\pdfoutput=1 

\usepackage{jheppub} 

\usepackage[T1]{fontenc} 
\usepackage{bm}

\newcommand{\be}{\begin{eqnarray}}
\newcommand{\ee}{\end{eqnarray}}
\newcommand{\ba}{\begin{array}}
\newcommand{\ea}{\end{array}}

\newcommand{\bea}{\begin{eqnarray}}
\newcommand{\eea}{\end{eqnarray}}

\newcommand{\bi}{\begin{itemize}}
\newcommand{\ei}{\end{itemize}}

\newcommand{\nn}{\nonumber}

\begin{document}

\title{Exact summation of  leading infrared logarithms in 2D effective field theories}

\author[a]{Jonas Linzen,}
\author[a,b]{Maxim V.~Polyakov,}
\author[b,c]{Kirill M. Semenov-Tian-Shansky,}
\author[b,d]{\\ Nika S.  Sokolova}


\affiliation[a]{Ruhr University Bochum, Faculty of Physics and Astronomy,
Institute of Theoretical Physics II, D-44780 Bochum, Germany}
\affiliation[b]{National Research Centre ``Kurchatov Institute'': Petersburg Nuclear Physics
Institute,
		 RU-188300 Gatchina, Russia}
\affiliation[c]{Saint Petersburg National Research Academic University
of the Russian Academy of Sciences,
RU-194021 St.~Petersburg, Russia}		
\affiliation[d]{St.~Petersburg State University, Faculty of Physics, Ul'yanovskaya ul.~3, RU-198504, Peterhof, St.~Petersburg, Russia}


\abstract{A method of exact all-order summation of leading infrared logarithms in two dimensional massless $\Phi^4$-type
non-renormalizable effective field theories (EFTs) is developed.
The method is applied to the
${\rm O}(N)$-symmetric
EFT, which is a two-dimensional sibling of the four dimensional
${\rm O}(N+1)/{\rm O}(N)$
sigma-model. For the first time the exact all-order summation of the
$\left(E^{2} \ln(1/E)\right)^n$
contributions (chiral logarithms) for the
$2 \to 2$
scattering amplitudes is performed in closed analytical form. The cases when the resulting amplitudes turn to be  meromorphic functions with an infinite number of poles (Landau poles) are identified. This provides the first explicit example of quasi-renormalizable field theories.

}

\maketitle

\section{Introduction}

The perturbative expansion for observables in Quantum Field Theory (QFT)
usually involves powers of  large logarithms of energy scale parameters.
This signals the necessity for rearrangement of the perturbation theory series
and consistent summation of large logarithmic contributions.
In  renormalizable QFTs this summation  is performed with the help
of the Renormalization Group (RG) equations. In particular, to sum up
the so-called leading logarithms (LLs)
(defined as the highest possible power of a large logarithm at a given loop order)
it suffices to take into account
the result of a one-loop calculation.

Taming large logarithmic corrections
in non-renormalizable Effective Field Theories (EFTs) turns to be a more difficult task.
In refs.~\cite{Kivel:2008mf,Koschinski:2010mr}
a generalization of the RG-approach  for generic massless
$\Phi^4$-type EFTs was constructed.
The corresponding equations have the
form of non-linear recurrence relations between the
leading large logarithm coefficients.
This allows to calculate the coefficients in front of
leading large logarithms to an arbitrary high loop order.
Let us illustrate this on the example of the
${\rm O}(N+1)/{\rm O}(N)$
non-linear $\sigma$-model in
$4D$
space-time dimensions defined by the action:
\be
S&=&\int d^4 x\ \frac12 g_{ab} (\Phi)\ \partial_\mu \Phi^a \partial^\mu \Phi^b \nn \\ & =&
\int d^4 x\ \left(\frac 12 \partial_\mu \Phi^a \partial^\mu \Phi^a +\frac{1}{2 F^2} (\Phi^a \partial_\mu\Phi^a)(\Phi^b \partial^\mu\Phi^b)
+ O\left(\Phi^6 \right)\right),
\label{eq:ON}
\ee
where
$\Phi^a$
is the
$N$-component scalar field,  $g_{ab}(\Phi)$ is a metric on the $S^N$-sphere of radius $F$ which has the dimension of the mass.
The non-renormalizable EFT
(\ref{eq:ON})
describes the interaction
of the Goldstone bosons in
$4D$.
The contribution of the leading infra-red%
\footnote{The term ``infrared'' refers to the low energy behavior: $E^2 \ll \mu^2$.}
logarithms to the low energy expansion of
an observable ${\cal M}$ ({\it e.g.}
the
$2 \to 2$
scattering amplitude) generically has the form:
\be
\label{eq:LLamp}
{\cal M}(E) =\frac{E^2}{F^2} \sum_{n=1}^\infty \omega_n \left[\frac{E^2}{(4\pi F)^2} \ln\left(\frac{\mu^2}{E^2} \right) \right]^{n-1}.
\ee
Here $E$ is the appropriate energy variable
($E= \sqrt{s}$ for the case of the $2 \to 2$
scattering amplitude)
and
$\mu$
is an arbitrary (in the LL-approximation) characteristic
energy scale, which can be seen as an analogue of the scale parameter of the running coupling constant of a renormalizable QFT.
The index
$n$
corresponds to the
$(n-1)$-th loop contribution in the theory
(\ref{eq:ON}).
The LL-coefficients
$\omega_n$
are related to each other by the non-linear
recurrence relations derived in refs.~\cite{Kivel:2008mf,Koschinski:2010mr}.
The summation of the LL-series in
Eq.~(\ref{eq:LLamp})
can be formally written as:
\be
{\cal M}(E)=\frac{E^2}{F^2} \ \Omega\left(\frac{E^2}{(4\pi F)^2} \ln\left(\frac{\mu^2}{E^2} \right)  \right),
\ee
in terms of the generating function for the LL-coefficients:
\be
\Omega(z) = \sum_{n=1}^\infty \omega_n z^{n-1}.
\ee
The function
$\Omega(z)$
encodes valuable information on the corresponding EFT in the non-perturbative regime.
For example, the position of the nearest to the origin singularity of the function
$\Omega(z)$
in the complex  $z$-plane
can provide us with exact exponents for the power behavior of various correlation
functions in configuration space in non-perturbative regime of the EFT, see
{\it e.g.}
refs.~\cite{Kivel:2008ry,Kivel:2009az,Perevalova:2011qi,Ananthanarayan:2018kly}.
Moreover, the function
$\Omega(z)$
may contain the information on the non-perturbative spectrum of masses
of a non-renormalizable EFT.

For the
$4D$
massless EFTs the explicit expressions for the function
$\Omega(z)$
was obtained only for the case of the large-$N$
limit in the
${\rm O}(N+1)/{\rm O}(N)$
\cite{Kivel:2008mf,Koschinski:2010mr}
and
${\rm SU}(N)/({\rm SU}(N-M)\times {\rm SU}(M))$
\cite{Polyakov:2010pt}
$\sigma$-models.
For the
${\rm O}(N+1)/{\rm O}(N)$
$\sigma$-model in the limit
$N\to\infty$
the exact result for the generating function
$\Omega(z)$
(corresponding to the isospin zero, $S$-wave scattering amplitude)
has the form:
\be
\Omega^{N\to\infty}(z)=\frac{N}{1-N z}.
\ee
The resulting amplitude contains a single Landau pole typical for the solution of
RG-equations in renormalizable QFTs. It is not surprising since in the large-$N$ limit
the theory
(\ref{eq:ON})
with
$F\sim \sqrt{N}$
is equivalent to a renormalizable QFT
(see {\it e.g.} general discussion in ref.~\cite{ANV}) and the amplitude
$\Omega(z)$
can be obtained with the help of RG-equations.

For more complicated theories (not admitting the description by means of the
RG-equations in the large-$N$ limit),
like the principal chiral field in $4D$, the exact form and analytical properties of
$\Omega(z)$
still remains unknown.
The numerical studies of the $4D$
$SU(\infty)\times SU(\infty)$
principal chiral field
\cite{Julia} indicate that, rather than having a single Landau pole,
$\Omega(z)$
can possess a much more complicated analytical structure.
Recently, in
ref.~\cite{QRI}
the so-called {\it quasi-renormalizable} (QR) QFTs were introduced.
In such QFTs the generating function of the LL-coefficients
$\Omega(z)$
is supposed to be meromorphic on a certain domain $\cal D$ in the complex-$z$ plane.
Additional singularities ({\it e.g.} branch 
points)
may also be present outside the domain
$\cal D$.
In a situation that we believe to be theoretically very interesting
$\Omega(z)$ is meromorphic on the whole complex-$z$ plane
and possesses an infinite number of poles, which can be
interpreted as a manifestation of an infinite number of the Landau poles in the theory.

The main goal of the present paper is to provide physical examples of
such QFTs.
We consider the
O$(N)$-symmetric two dimensional massless
$\Phi^4$-type EFT defined by the following action:
\be
&&
\hspace{-0.4cm} S= \int d^2 x \left(\frac{1}{2} \partial_\mu {\Phi^a}  \partial^\mu {\Phi^a} -
g_1 (\partial_\mu {\Phi^a}  \partial^\mu {\Phi^a})
(\partial_\nu {\Phi^b}  \partial^\nu {\Phi^b})-
g_2 (\partial_\mu {\Phi^a}  \partial_\nu {\Phi^a})
(\partial^\mu {\Phi^b}  \partial^\nu {\Phi^b})\right). \nn \\ &&
\label{LagrangianBissextile}
\ee
Here
${\Phi^a}(x)$, $a=1,\ldots,N$
is the
$N$-component vector in the internal symmetry space;
$g_{1,\,2}$
are the coupling constants at the two possible vertices involving
$4$ derivatives. The dimension of
$g_{1,\,2}$
is
$-2$
in the mass units.

The {\it non-renormalizable} EFT
(\ref{LagrangianBissextile})
can be seen as a two-dimensional sibling of the four-dimensional
EFT (\ref{eq:ON}) -- it has the same  symmetry and the same structure of the  leading logarithms
(\ref{eq:LLamp}) as in 4D $\sigma$-model (\ref{eq:ON}).
We name the theories of the type (\ref{LagrangianBissextile}) as
{\it bi-quartic, theories}%
\footnote{In Ref.~\cite{QRI} the term {\it bissextile} was used instead.} since
the interaction part of the corresponding
Lagrangian is built of four-point interactions with
four derivatives. 
Below we argue that this property of the interaction leads to the appearance
of an  infinite number of poles in the LL-approximation for the binary scattering amplitude, thus providing  explicit examples for quasi-renormalizable QFTs.

 A few words of comments are necessary here on the manifestation of the infrared divergences
in the theory (\ref{LagrangianBissextile}). Indeed the infrared divergences are
known to plague severely the perturbative calculations in massless theories in 2D.
This issue is discussed in details {\it e.g.} in Refs.~\cite{Brezin:1976qa,Brezin:1976ap,David:1980rr}
in connection with the calculations in the $2D$ non-linear sigma model.
We would like to stress that infrared divergences represent a minor issue
for the calculation of leading infrared logarithmic corrections in the  bi-quartic theory in 2D.  As usual, the loop calculations require introduction
of a small regulating mass $m$ for the field $\Phi$ in (\ref{LagrangianBissextile}).
However, the leading infrared logarithmic approximation%
 \footnote{Let us stress that within the leading infrared log approximation for
 $2 \to 2$ scattering amplitude
 our goal is to sum up the powers of logarithms of energy variable of the form $ \left( \log \frac{1}{s} \right)^{n-1}$, where $n$ refers to ``number of loops $+1$''. At this point
 there is a distinction with the common RG problem of  summing up
  $ \left( \log \mu^2 \right)^{n-1}$ contributions in renormalizable theories.} for the $2 \to 2$ in the bi-quartic theory in 2D turns to be infrared finite and the massless limit can be
taken  safely. This comes from the fact that the interaction in (\ref{LagrangianBissextile})
is too soft to produce power-like singularities in the regulating mass. On the other
hand, the logarithmic singularity in the regulating mass can not show up in our
LL-approximation since we sum up terms with maximal power of $\log \frac{1}{s}$ from each loop order. A $\log \frac{1}{m^2}$ entry will reduce the overall power of $\log \frac{1}{s}$ and
hence the corresponding term will not contribute to the LL-accuracy.

\section{Scattering amplitudes of the O$(N)$-symmetric bi-quartic theory in the leading logarithmic approximation}

In this section we consider the recurrence relations for the LL-coefficients of
the
$2 \to 2$
(binary) scattering amplitude in the O$(N)$-symmetric bi-quartic theory
(\ref{LagrangianBissextile})
in $2D$.
We refer the reader to Appendix \ref{AppA} for the derivation of
the recurrence relations for the case of a generic $\Phi^4$-type massless EFT in $2D$.
Here we explicitly present the details specific for the theory
(\ref{LagrangianBissextile}).

The amplitude of the binary scattering process
 \be
 \Phi_a(p_1)+\Phi_b(p_2)\to \Phi_c(p_3)+\Phi_d(p_4)
 \label{2to2scattering1}
 \ee
in the ${\rm O}(N)$-symmetric bi-quartic theory
(\ref{LagrangianBissextile})
admits the following decomposition in the irreducible representations of the
${\rm O}(N)$ group with respect to the $s$-channel isospin:
\be
&&
 {\cal M}_{abcd}(s,t,u)
=  \sum_{I=0}^2 P^{I}_{abcd} {\cal M}^{I}(s,t,u).
\label{BinaryAmpl}
\ee
Here
${\cal M}^{I}(s,t,u)$
are the invariant amplitudes,
$s$, $t$
and
$u$
are the usual Mandelstam variables:
\be
s=(p_1+p_2)^2; \ \ t=(p_1-p_4)^2; \ \ u=(p_1-p_3)^2. \nonumber
\ee
The projection operators on the invariant subspaces read:
\be
&&
P^{I=0}_{abcd}=\frac{1}{N} \delta_{ab}\delta_{dc}; \ \ \
P^{I=1}_{abcd}=\frac{1}{2}  \left(
\delta_{ad}\delta_{bc}-\delta_{ac}\delta_{bd}
\right); \nn \\ &&
P^{I=2}_{abcd}=\frac{1}{2}
\left(
\delta_{ad}\delta_{bc}+\delta_{ac}\delta_{bd}
\right)-\frac{1}{N} \delta_{ab}\delta_{cd}.
\label{Isospin_projectors}
\ee
The projection operators
(\ref{Isospin_projectors})
satisfy the completeness relation:
\be
P^{I=0}_{abcd}+P^{I=1}_{abcd}+P^{I=2}_{abcd}=\delta_{ad}\delta_{bc}.
\label{completeness_rel}
\ee

The crossing relations between the invariant amplitudes
${\cal M}^{I}(s,t,u)$
can be written in terms of the isospin crossing matrices:
\be
{\cal M}^{I}(s,t,u)=C_{st}^{IJ} {\cal M}^{J}(t,s,u), \nn \\
{\cal M}^{I}(s,t,u)=C_{su}^{IJ} {\cal M}^{J}(u,t,s), \\
{\cal M}^{I}(s,t,u)=C_{tu}^{IJ} {\cal M}^{J}(s,u,t).\nn
\ee
The crossing matrices  are expressed  through the
invariant projection operators
(\ref{Isospin_projectors})
as
\be
C_{su}^{IJ }= \frac{1}{d_I} P^I_{abcd} P^{J }_{bdac}; \ \ \
C_{st}^{IJ }= \frac{1}{d_I} P^I_{abcd} P^{J }_{cbad}; \ \ \
C_{tu}^{IJ }= \frac{1}{d_I} P^I_{abcd} P^{J }_{bacd},
\ee
where
\be
d_I=P^I_{abba}= \Big\{1,\, \frac{N(N-1)}{2}, \, \frac{(N+2)(N-1)}{2} \Big\}
\label{DimSubS}
\ee
stand for the dimensions of the corresponding irreducible representations.
The explicit expressions for the crossing matrices consistent with the
completeness relation
(\ref{completeness_rel})
read
\be
&&
C_{su}=\left(
\begin{array}{ccc}
 \frac{1}{N} & \frac{1-N}{2} &
   \frac{N^2+N-2}{2 N} \\
 -\frac{1}{N} & \frac{1}{2} & \frac{N+2}{2
   N} \\
 \frac{1}{N} & \frac{1}{2} & \frac{N-2}{2
   N}
\end{array}
\right); \ \ \
C_{st}=
\left(
\begin{array}{ccc}
 \frac{1}{N} & \frac{N-1}{2} &
   \frac{N^2+N-2}{2 N} \\
 \frac{1}{N} & \frac{1}{2} &
   -\frac{N+2}{2 N} \\
 \frac{1}{N} & -\frac{1}{2} &
   \frac{N-2}{2 N}
\end{array}
\right); \nn \\ &&
C_{tu}=
\left(
\begin{array}{ccc}
 1 & 0 &  0 \\
 0 & -1 & 0 \\
 0 & 0 & 1
\end{array}
\right).
\label{CrossM}
\ee

In two space-time dimensions the scattering amplitudes
${\cal M}^{I}(s,t,u)$
can be further decomposed into the transmission and reflection
amplitudes
${\cal M}^{I, \; \{ T, \,R\}}(s)$
which depend only on the invariant total energy variable
$s$
(see Appendix~\ref{AppA} for definitions and discussion).

To the tree-level accuracy,
the expressions for the transmission and
reflection isotopic invariant amplitudes
${\cal M}^{I, \; \{ T, \,R\}}(s)$
in the theory
(\ref{LagrangianBissextile})
read:
\be
&&
{\cal M}^{I=0, \; T}(s)={\cal M}^{I=0, \; R}(s)=s^2(2g_1(N+1)+g_2(N+3)), \nn \\ &&
{\cal M}^{I=1, \; T}(s)=-{\cal M}^{I=1, \; R}(s)=s^2(g_2-2g_1),  \label{Tree-level_amp} \\ &&
{\cal M}^{I=2, \; T}(s)={\cal M}^{I=2, \; R}(s)=s^2(2g_1+3g_2).\nn
\ee

The leading logarithmic contributions
into the transition and reflection invariant amplitudes
we would like to compute admit
the following parametrization:
\be
\label{eq:MLLA}
{\cal M}^{I, \; \{ T, \, R \}}\Big|_{\text{LL}}(s)= s^2 \sum_{n=1}^\infty
\omega^{I, \; \{ T, \, R \}}_n\  \left[ \frac{s}{4\pi}  \ln \left( \frac{\mu^2}{s} \right) \right]^{n-1},
\ee
where we introduced transmission and reflection
LL-coefficients
$\omega^{I, \; \{ T, \, R \}}_n$
for given isospin
$I$. Note that the
index $n$ refers to ``number of loops plus $1$''. The tree-level LL-coefficients can
be directly read off the tree-level amplitudes
(\ref{Tree-level_amp}):
\be
&&
\omega_1^{I=0, \; T} =\omega_1^{I=0, \; R} = (2g_1(N+1)+g_2(N+3)); \nn \\ &&
\omega_1^{I=1, \; T} =-\omega_1^{I=1, \; R} = (g_2-2g_1);  \nn \\ &&
\omega_1^{I=2, \; T} =\omega_1^{I=2, \; R} = (2g_1+3g_2).
\label{tree_level_LL}
\ee

The explicit form of the system of non-linear recurrence relations for
the LL-coefficients (see Appendix \ref{AppA} for the derivation)
reads%
\footnote{Note that there is no summation over
the repeating $J$ in terms like
$\omega_{i}^{J, \; \{T,\,R\}} \cdot \omega_{n-i}^{J, \; \{T,\,R\}}$
in (\ref{System_RR}):
the two $J$ entries should be understood here as a sole summation index in the
convolution with the isospin crossing matrices.}:
\be
&&
\omega_{n}^{I, \; T}= \frac{1}{2(n-1)}
\sum_{k=1}^{n-1} \sum_{J=0}^2
\left( \delta^{IJ}-(-1)^n C_{su}^{IJ} \right)
\left(
\omega_{k}^{J, \; T} \cdot \omega_{n-k}^{J, \; T}+
\omega_{k}^{J, \; R} \cdot \omega_{n-k}^{J, \; R}
\right); \nn \\ &&
%
\omega_{n}^{I, \; R}= \frac{1}{2(n-1)}  \sum_{k=1}^{n-1}
\sum_{J=0}^2
\left(\delta^{IJ}-(-1)^n C_{st}^{IJ} \right)
\left(
\omega_{k}^{J, \; T} \omega_{n-k}^{J, \; R}+
\omega_{k}^{J, \; R} \omega_{n-k}^{J, \; T}
\right).
\label{System_RR}
\ee
Here the isospin crossing matrices
$C_{su}$
and
$C_{st}$
are defined in
(\ref{CrossM}).
The initial conditions for the recurrence system are provided
by the tree-level result
(\ref{tree_level_LL}).
The transmission LL-coefficients up to four-loops obtained from the system of recurrence equations
(\ref{System_RR})
are listed in Table~\ref{LLI=0}.

\begin{table}[h]
\centering
\begin{tabular}{lccc}
\hline
\hline
$n$ & $I=0$ & $I=1$ & $I=2$ \\
\hline
$1$ & $(2g_1(N+1)+g_2(N+3)) F^2$ & $(-2g_1+g_2) F^2$ & $(2g_1+3g_2) F^2$ \\
$2$ & $(N-1)(N+2)$ & $(N+2)$ & $-(N+2)$ \\
$3$ & $(N-1)(N+2)^2$ & $-(N+2)^2$ & $(N-2)(N+2)$ \\
$4$ & $\frac{1}{3} (N-1)(N+2)^2(3N-2)$ & $\frac{1}{3}(N+2)^2(3N-2)$ & $-\frac{1}{3}(N+2)^2(3N-2)$ \\
$5$ & $\frac{1}{3}(N-1)(N+2)^3(3N-1)$ & $-\frac{1}{3}(N+2)^3(3N-1)$ & $\frac{1}{3}(N+2)^2(N-2)(3N-1)$ \\
\hline
\hline
\end{tabular}
\caption{\label{LLI=0}  Table of $I=0, \,1,\,2$ transmission LL-coefficients (multiplied by $F^{2n}$) for the
bi-quartic model (\ref{LagrangianBissextile}) up to four loops. Here $1/F^2=2 g_1+g_2$ is the combination of coupling constants which enters the LL-approximation.  }
\end{table}

It is straightforward to check that for all $n$
\be
\omega_n^{I, \; T} = (-1)^I \omega_n^{I, \; R},
\ee
that is just a consequence of the Bose symmetry.
Moreover it turns our that for
$n \ge 2$
\be
\label{rat1}
\frac{\omega_{n}^{I=1, \; T}}{\omega_{n}^{I=0, \; T}}=
\begin{cases}
\frac{1}{N-1}, \ \ \ n -\text{even}; \\
-\frac{1}{N-1}, \ \ \ n -\text{odd};
\end{cases}
\ee
and
\be
\label{rat2}
\frac{\omega_{n}^{I=2, \; T}}{\omega_{n}^{I=0, \; T}}=
\begin{cases}
-\frac{1}{N-1}, \ \ \ n -\text{even}; \\
\frac{N-2}{(N-1) (N+2)}, \ \ \ n -\text{odd}.
\end{cases}
\ee
Therefore, to solve the recurrence system
(\ref{System_RR})
it suffices to
consider just the
$I=0$ transmission LL-coefficients
$\omega_{n}^{I=0, \; T}$.
It is curious that for
$n \ge 2$
these coefficients depend only on the particular combination of
the coupling constants of (\ref{LagrangianBissextile}):
\be
\frac{1}{F^2} \equiv 2g_1+g_2.
\label{Gdef}
\ee
Here the coupling constant $F$ has the dimension of mass.

With the help of eqs.~(\ref{rat1}), (\ref{rat2})
one can show that the non-linear recurrent system
(\ref{System_RR})
for
$\omega_{n}^{I=0, \; T}$
is equivalent to the following non-linear recurrent relation:
\be
f_n=\frac{1}{n-1} \sum_{k=1}^{n-1}
\left(
A_0+ (-1)^n A_1 +  (-1)^k A_2
\right)  f_k f_{n-k}; \ \ \ f_1=1,
\label{recrel_master1}
\ee
where the coefficients
$A_i$
read:
\be
A_0= 1 + \frac{1}{(N+2)(N-1)}; \ \  A_1=-\frac{N+1}{(N+2)(N-1)};
\ \  A_2=-\frac{2}{(N+2)(N-1)}.
\ee
For
$n\ge 2$
the
$I=0$
transmission LL-coefficients
$\omega_{n}^{I=0, \; T}$
are expressed through
the solution of
(\ref{recrel_master1})
as
\be
\label{omegaomicron}
\omega_{n}^{I=0, \; T}=f_n \times  \left( \frac{(N+2)(N-1)}{N F^2} \right)^n,
\ee
where $F$ is introduced in (\ref{Gdef}).

Some general properties of the recurrence equations of the type (\ref{recrel_master1}) were recently discussed  in
\cite{QRI} in the context
of the quasi-renormalizable field theories.
The main distinctive feature of the QR field theories is that the ``running coupling constant'' (defined
through the $2 \to 2$ particle scattering amplitude in the LL-approximation)
possesses a  more complicated analytical structure, rather than a simple Landau pole.

For the case
$A_1=A_2=0$
the recurrence relation
eq.~(\ref{recrel_master1})
reduces to the form similar to that of the usual renormalizable QFT.
The corresponding solution 
contains a single Landau pole with position defined by the
coefficient
$A_0$.
The solutions for more general cases exhibit much more
complicated analytical structure
\cite{QRI}.

As we discussed above,
all possible invariant binary scattering amplitudes
in the bi-quartic model
(\ref{LagrangianBissextile})
can be expressed in terms of a single amplitude
${\cal M}^{I=0, \; T}(s)$.
We define
the dimensionless amplitude (generating function) in the LL-approximation as:
\be
\label{eq:Omdef}
\Omega(z)=\sum_{n=2}^\infty \omega_{n}^{I=0, \; T} z^{n-1}.
\ee
In what follows, we call the function
$\Omega(z)$
(\ref{eq:Omdef})
as the amplitude in the LL-approximation, or LL-amplitude for brevity.
All possible binary scattering amplitudes
in the bi-quartic model
(\ref{LagrangianBissextile})
in the LL-approximation can be expressed through
$\Omega(z)$
using the relations:
\be
{\cal M}^{I=0, \; T}
\Big|_{\text{LL}}
(s)&=&s^2\left[2g_1(N+1)+g_2(N+3)\right]+\frac{s^2}{F^2}\ \Omega\left( \frac{s}{2\pi F^2} \ln\left(\frac{\mu^2}{s}\right)\right); \nn \\
\label{eq:ampVSom}
{\cal M}^{I=1, \; T}
\Big|_{\text{LL}}
(s)&=&s^2(g_2-2g_1)-\frac{s^2}{(N-1)F^2}\ \Omega\left( -\frac{s}{2 \pi F^2} \ln\left(\frac{\mu^2}{s}\right)\right);  \\
{\cal M}^{I=2, \; T}
\Big|_{\text{LL}}
(s)&=&s^2(2g_1+3g_2)\nn \\
&-&\frac{2s^2}{(N+2)(N-1)F^2}\
\left[\Omega\left(\frac{s}{2 \pi F^2} \ln\left(\frac{\mu^2}{s}\right)\right)
-\frac{N}{2}\Omega\left( -\frac{s}{2 \pi F^2} \ln\left(\frac{\mu^2}{s}\right)\right)\right]; \nn\\
{\cal M}^{I, \; R}\Big|_{\text{LL}}(s)&=&(-1)^I\ {\cal M}^{I, \; T}\Big|_{\text{LL}}(s).\nn
\ee
It worths mentioning the special case $N=1$. It corresponds to surviving  of just $I=0$
invariant subspace ({\it cf.} eq.~(\ref{DimSubS}) for the dimensions of the invariant subspaces). As it can be seen for Table~\ref{LLI=0}, starting from the one-loop order the LL-coefficients
$\omega_{n}^{I=0, \; T,\,R}$ vanish and the solution for the LL-amplitude (\ref{eq:Omdef}) is just $\Omega(z)=0$. However, as pointed out in Sec.~\ref{EllipticFsol}, one can rigourously defile the limit $\lim_{N \to 1} \frac{\Omega(z)}{N-1}$ that can be expressed in terms
of the  Weierstra{\ss} elliptic function.

In the next sections we present the non-trivial solutions for the LL-amplitude (\ref{eq:Omdef}).
We also note that using the technique of
refs.~\cite{Kivel:2009az}, \cite{Ananthanarayan:2018kly}
one can compute the LL-approximation
for various form factors
as well as the correlation functions of two currents
in the bi-quartic theory
(\ref{LagrangianBissextile})
in terms of the LL-amplitude
(\ref{eq:Omdef}).

\section{Generalizing RG-equations: non-linear differential equations for the LL-amplitude }

In order to compute the LL-amplitude
(\ref{eq:Omdef})
in the bi-quartic theory
(\ref{LagrangianBissextile}) we need to solve the
recurrence equation
(\ref{recrel_master1}).
To reduce the recurrence equation
(\ref{recrel_master1})
to a differential equation, we introduce the generating function
\be
f(z)=\sum_{n=1}^\infty f_n z^{n-1}.
\ee
This function satisfies the following differential equation equivalent to (\ref{recrel_master1}):
\be
\label{eq:GRGW}
\frac{d}{dz} f(z) =A_0\  f(z)^2 +A_1\ f(-z)^2-A_2\  f(z) f(-z),\quad f(0)=1,
\ee
Note that for the case of a renormalizable QFT
$A_1=A_2=0$, the equation
(\ref{eq:GRGW})
is reduced to:
\be
\frac{d}{dz} f(z) =A_0\  f^2(z),
\ee
which has the form of the one-loop RG-equation with the solution:
\be
f(z)=\frac{1}{1-A_0 z}.
\ee
This solution corresponds to the famous Landau pole and leads to the classification of renormalizable QFTs  into
IR- or UV- asymptotically free theories, depending on the sign of
the coefficient $A_0$.

For the case of a non-renormalizable EFT we need to consider a much more complicated
non-linear differential equation
(\ref{eq:GRGW}).
To address this issue, it is convenient to introduce the even
($u(z)=u(-z)$)
and odd
($v(z)=-v(-z)$)
parts of the generating function $f(z)$:
\be
u(z)=\sum_{n=1 \atop \text{odd}}^\infty f_n z^{n-1}; \ \ \
v(z)=\sum_{n=2 \atop \text{even}}^\infty f_n z^{n-1}.
\ee
The recurrence relation
(\ref{recrel_master1})
is equivalent to the following system of non-linear differential equations for
$u(z)$ and $v(z)$:
\be
\begin{cases}
v'(z)= (A_0+A_1-A_2) u^2(z)+ (A_0+A_1+A_2) v^2(z);
\\
u'(z)=2 (A_0-A_1) u(z) v(z);
\end{cases}
\label{DifUr}
\ee
with the initial conditions
$$
u(0)=1; \ \ \ v(0)=0.
$$
We can use the second equation in
(\ref{DifUr})
to express $v(z)$ as:
\be
v(z)= -\frac{1}{2(A_0-A_1)} u(z) \frac{d}{dz} \frac{1}{u(z)}.
\label{g_expressed}
\ee
It is convenient to introduce the new variable
\be
l(z)= \frac{1}{u(z)}; \ \ \ l(0)=1; \ \ \ l'(0)=0.
\label{def_l(z)}
\ee
Taking into account
(\ref{g_expressed}),
the first equation in
(\ref{DifUr})
can then be rewritten as:
\be
l(z) l''(z)= \alpha_1 \left( l'(z) \right)^2+ \alpha_0,
\ee
where
\be
&&
\alpha_0=-2 (A_0-A_1) (A_0+A_1-A_2)= -\frac{2N^3}{(N-1)^2(N+2)};  \\ &&
\alpha_1=1-\frac{(A_0+A_1+A_2)}{2 (A_0- A_1)}= \frac{N+2}{2N}.
\ee
Now we make use of the standard trick employed for autonomous differential equations and
perform the substitution
\be
\nu(l)=l'(z); \ \ \ l''(z)=\nu \nu'; \ \ \text{where} \ \  \nu' \equiv \frac{d}{dl} \nu(l).
\ee
This produces the equation
\be
l \nu \nu'= \alpha_1 \nu^2+\alpha_0,
\ee
which can be integrated
and gives
\be
\alpha_1 \log(l C)= \frac{1}{2} \log \left(\alpha_1 \nu^2+\alpha_0 \right).
\ee
The integration constant $C$ is fixed from the $\nu(1)=0$ condition:
\be
\alpha_1 \log(C)=\frac{1}{2} \log \left( \alpha_0 \right).
\ee
Thus, we arrive to the following differential equation for
$l(z)$ (\ref{def_l(z)}):
\be
\alpha_1 \left[ l'(z)) \right]^2= \alpha_0 l^{2 \alpha_1}(z)- \alpha_0; \ \ \ l(0)=1,
\label{Ell_1}
\ee
that allows to express the even part of the generating function
$f(z)$.
The odd part of the generating function
$f(z)$
can be expressed through
(\ref{g_expressed}).
In the next section we consider several solutions of the differential equation
(\ref{Ell_1}).

\section{Solutions for the LL-scattering-amplitude }

The differential equation
(\ref{Ell_1})
can be formally viewed as the equation of motion of a  one dimensional mechanical system.
Indeed, introducing the ``time variable'':
\be
t= \frac{N}{(N-1)(N+2)} z,
\ee
and the ``coordinate''
\be
q(t)=l\left(\frac{(N-1)(N+2)}{N}\ t\right).
\ee
we obtain the equation of motion for a particle of the mass
$$
m=\frac{1}{2 N^2}
$$
and the total energy $E=1$ in one dimension (along the coordinate $q$):
\be
\label{eq:mech_system}
\frac{m\ \dot q(t)^2}{2}   +q(t)^\gamma  =1,
\ee
where
$$
\gamma=\frac{N+2}{N}
$$
is the exponent of the potential. The initial condition at
$t=0$ corresponds to the particle
resting at
$q=1$.
{Note that the parameters $m$ and $\gamma$ of the effective
mechanical system
(\ref{eq:mech_system})
are singular in the limit $N \to 0$.
We address this issue in Sec.~\ref{ElFsolutions}.}

Using the relation
(\ref{omegaomicron})
together with (\ref{def_l(z)}), (\ref{g_expressed})
one can express the  LL-amplitude
$\Omega(z)$
(\ref{eq:Omdef})
through the solution of the equation of motion
(\ref{eq:mech_system}):
\be
\label{eq:omS}
\Omega(z)=\frac{(N-1)(N+2)}{N}\ \left(\frac{1}{q(z)}-1\right)-\frac{N-1}{2 N}\ \frac{d}{dz} \ln(q(z)).
\ee
We conclude that the problem of summation of LLs for the binary scattering amplitudes
in the ${\rm O}(N)$-symmetric bi-quartic model
(\ref{LagrangianBissextile})
is reduced to the analysis of the one-dimensional
motion of the mechanical system
(\ref{eq:mech_system}).
The knowledge of the function
$\Omega(z)$
allows to compute all scattering amplitudes
in LL-approximation in the theory
(\ref{LagrangianBissextile})
with the help of eq.~(\ref{eq:ampVSom}).

We note that with the change of the coordinate
$q(t)=r(t)^{\frac{2}{2-\gamma}}$
one  obtains the mechanical system equivalent to
(\ref{eq:mech_system}):
\be
\label{eq:mech_system1}
\frac{M\ \dot{r}(t)^2}{2}+\left[2-r(t)^\delta\right]=1, 
\ee
where the corresponding ``mass'' and exponent of the ``potential'' are
given by
\be
M=\frac{2}{(N-2)^2}; \quad \delta=\frac{2\gamma}{\gamma-2}= - \frac{2(N+2)}{N-2}.
\ee
{ The parameters $M$ and $\delta$ of the equivalent mechanical system
(\ref{eq:mech_system1})
turn out to be singular for $N=2$. This is another special case we consider
in Sec.~\ref{ElFsolutions}.}

For the case of the exponents of the potentials
$\gamma$
and
$\delta$
equal to
$0,\,1,\, 2,\, 3,\, 4$
the mechanical systems
(\ref{eq:mech_system}), (\ref{eq:mech_system1})
can be solved explicitly in terms of elliptic functions (or their degeneracies).
The values of the corresponding theory parameters%
\footnote{The non-entire values of $N$ should be understood in a sense of analytical
continuation ({\it cf.} ref.~\cite{ANV}).}
 and a short summary of
solutions is presented in
Tables~\href{Table2}{2}, \href{Table3}{3}.
Below, in Secs.~\ref{ElFsolutions}, \ref{EllipticFsol},
we present a detailed analysis of the corresponding solutions.
A qualitative analysis of the general case is given in
Sec.~\ref{Qual_analys}.

\begin{table}
\label{Table2}
\centering
\begin{tabular}{|c|c|c|c|l|}
\hline
$N$& $\gamma$ & $m$ &  $\Omega(z)$ & Comments: \\
\hline
\hline
$-2$ &0 & $\frac{1}{8}$ & $0$& free motion of (\ref{eq:mech_system}); no LLs; \\
\hline
$N \to \infty$ &  $1$ & $m \to 0$ & eq.~(\ref{N_to_infty})& motion of (\ref{eq:mech_system}) under constant force;   \\
\hline
$2$ & $2$ & $\frac{1}{8}$ & eq.~(\ref{N=2}) & harmonic oscillator case;  trigonometric solution;\\
\hline
$N \to 1$ & $3$ & $\frac{1}{2}$ & eq.~(\ref{eq:Bacher}) & elliptic solution; \\
\hline
$\frac{2}{3}$ & $4$ & $\frac{9}{8}$ & eq.~(\ref{Elliptic_N=2/3}) & elliptic solution; \\
\hline
\end{tabular}
\caption{Summary of solutions for the mechanical system (\ref{eq:mech_system}) expressed
in terms of elliptic functions (and their degeneracies).}
\end{table}

\begin{table}[tbp]
\centering
\begin{tabular}{|c|c|c|c|l|}
\hline
$N$& $\delta$ & $M$ & $\Omega(z)$ & Comments: \\
\hline
\hline
$-2$  & $0$ & $\frac{1}{8}$ & $0$ &   $\gamma=0$ case; see Table~\href{Table2}{2}.\\
\hline
$-\frac{2}{3}$  &$1$ & $\frac{9}{32}$ & eq.~(\ref{N=minus2/3})& motion of (\ref{eq:mech_system1}) under constant force; \\
\hline
$N \to 0$  &$2$ & $\frac{1}{2}$ & eq.~(\ref{N_to_0})& motion of (\ref{eq:mech_system1}) in inverted
harmonic potential; $\; \; \; \; \; \; \; \; \;$\\
\hline
$\frac{2}{5}$  &$3$ & $\frac{25}{32}$ & eq.~(\ref{Elliptic_N=2/5})& elliptic solution; \\
\hline
$\frac{2}{3}$  &$4$ & $\frac{9}{8}$ & eq.~(\ref{Elliptic_N=2/3})&  $\gamma=4$ case; see Table~\href{Table2}{2}.\\
\hline
\end{tabular}
\label{Table3}
\caption{Summary of solutions for the mechanical system (\ref{eq:mech_system1}) expressed
in terms of elliptic functions (and their degeneracies).}
\end{table}

\subsection{Solutions in terms of elementary functions}
\label{ElFsolutions}

For the cases
$\gamma=0,1,2$
the motion  of  the mechanical system
(\ref{eq:mech_system}) is described
in terms of elementary functions.
These cases correspond to the ``free particle'', ``motion under a constant force'' and a ``harmonic oscillator'' respectively.
Another elementary
solution occurs for $\gamma=-2$, that corresponds to
the ``motion under the constant force'' of the  system (\ref{eq:mech_system1}),
and for
$\gamma\to\infty$,
that corresponds to the ``inverted harmonic potential'' in eq.~(\ref{eq:mech_system1}).

\vspace{0.2cm}
\bi
\item
\noindent
\underline{ $N=-2$, $\gamma=0$, $m=\frac{1}{8}$}\\
This case corresponds to the ``free particle motion'' of the mechanical system eq.~(\ref{eq:mech_system}).
The solution of eq.~(\ref{eq:mech_system}) is trivial: $q(t)=1$.
This leads to
$$\Omega(z)=0.$$
Therefore, for $N=-2$ no contribution of leading logarithms for $n>1$
come in the bi-quartic model
(\ref{LagrangianBissextile}).
\vspace{0.2cm}

\item
\noindent
\underline{  $N\to\infty$,  $\gamma=1$, $m\to 0$}\\
This case corresponds to the ``motion under a constant force''.
Note that in this case the mass of the particle tends to zero $m\sim 1/(2 N^2)$, therefore we rescale the time variable $t\to N t$ in eq.~(\ref{eq:mech_system}). The resulting
solution corresponds to the ``motion with uniform acceleration'' to the left from the initial position $q=1$:
\be
q(t)=1- (N t)^2.
\ee
With the help of eqs.~(\ref{eq:omS}) we obtain the large-$N$ solution for the
LL-amplitude in the bi-quartic model
(\ref{LagrangianBissextile}):
\be
\Omega(z) =\frac{N}{1-Nz}-N.
\label{N_to_infty}
\ee
We conclude that in this case the LL-amplitude possesses a single Landau pole. This is a typical result for the summation of LLs by means of the RG-equation.
The result
(\ref{N_to_infty})
is not surprising since in limit
$N\to\infty$
(assuming the coupling constants
$g_i\sim 1/N$ for $N \to \infty$)
the theory
(\ref{LagrangianBissextile})
is equivalent to a renormalizable field theory, see
{\it e.g.}
discussion in \cite{ANV}.

\item
\noindent
\underline{  $N=-\frac{2}{3}$, $\delta=1$, $M=\frac{9}{32}$}\\
This case corresponds to the ``motion under the action of the constant force'', see eq.~(\ref{eq:mech_system1}):
\be
r(t) =1+\frac{16}{9} t^2.
\ee
Correspondingly, the function $q(t)$ is expressed as:
\be
q(t)=\sqrt{r(t)}=\sqrt{1+\frac{16}{9} t^2}.
\ee
Finally, with the help of eq.~(\ref{eq:omS}) we obtain the solution for the LL scattering amplitude in the bi-quartic theory (\ref{LagrangianBissextile}) for $N=-\frac{2}{3}$:
\be
\Omega(z) = 10 \left(\frac{1}{\sqrt{9+16 z^2}}-\frac{2z}{9+16 z^2}\right)-\frac{10}{3}.
\label{N=minus2/3}
\ee
This is an example of a QFT with the LL-amplitude which has a more complicated analytical structure than that obtained with help of a usual leading order RG-equation for a renormalizable QFT.
It contains two branch 
points and poles at $z=\pm  \frac{3}{4}i$.

\item
\noindent
\underline{$N=2 $,  $\gamma=2$, $m=\frac{1}{8}$ }\\
This case corresponds to the ``harmonic oscillator''. The solution is the text-book periodic harmonic motion:
\be
q(t)=\cos(4 t).
\ee
This leads to the following LL-scattering-amplitude in the bi-quartic theory (\ref{LagrangianBissextile})
for $N=2$:
\be
\Omega(z) =\frac{2}{\cos(4z)}+\tan(4 z)-2.
\label{N=2}
\ee
This is a remarkable result. We conclude that for
$N=2$
the bi-quartic theory
(\ref{LagrangianBissextile})
exhibits an infinite number of the Landau poles on the real axis. The positions and residues of these poles are:
\be
z_k^{(1)}&=&\frac{\pi}{8} (4 k+1),\ k\in \mathbb{Z}, \quad \underset{\  \ z=z_k^{(1)}}{\rm Res} \Omega(z)=-\frac 34; \\
z_k^{(2)}&=&\frac{\pi}{8} (4 k+3),\ k\in \mathbb{Z}, \quad \underset{\  \ z=z_k^{(2)}}{\rm Res} \Omega(z)=\frac 14.
\ee
This case provides a live physical realization of  quasi-renormalizable QFTs introduced in ref.~\cite{QRI}. The presence of an infinite number of
poles in the LL-amplitude
(\ref{N=2})
may hint on the non-trivial mass spectrum of the theory
(\ref{LagrangianBissextile})
for
$N=2$.

{ We also note that $N=2$ turns out to be the special case from the point of view
of the equivalent mechanical system (\ref{eq:mech_system1}) as the corresponding parameters $M$ and $\delta$
become singular. However, we verified explicitly that (\ref{N=2}) provides the solution for the initial recurrent system (\ref{System_RR}). The solution (\ref{N=2}) is regular at $z=0$
hence being analytic functions in some circle around the point $z=0$.
Inside this circle the solution is unique as its Taylor expansion coincides with that
generated by the initial recurrence system (2.12).}
\item
\noindent
\underline{$N\to0 $,  $\delta=2$, $M=\frac{1}{2}$ }\\
This case  corresponds to the ``motion in an inverted harmonic potential'' (see eq.~(\ref{eq:mech_system1})) and again  has the text-book solution:
\be
r(t)=  \cosh(2 t).
\ee
Correspondingly, the function $q(t)$ is expressed as:
\be
q(t)=\frac{1}{r(t)^N}=\frac{1}{ \cosh(2 t)^N},
\ee
where we keep only the leading order terms in the
$N\to 0$
limit.
This solution leads to the following LL-scattering-amplitude in the bi-quartic theory (\ref{LagrangianBissextile}) in the limit
$N\to 0$:
\be
\Omega(z)=-\ln[\cosh(2 z)]-\tanh(2 z).
\label{N_to_0}
\ee
Therefore, the theory
(\ref{LagrangianBissextile})
in the limit of vanishing number of components
($N\to 0$) provides another example of a quasi-renormalizable QFT.
In this case the scattering amplitude has an infinite set of pole at:
\be
z_k=i \frac{\pi}{4}\left(2 k+1\right),\ k\in \mathbb{Z}, \quad \underset{\  \ z=z_k}{\rm Res} \Omega(z)=-\frac 12.
\ee
\ei

{
The case $N \to 0$ turns out to be singular from the point of view of the
equivalent mechanical system (\ref{eq:mech_system}).
We have checked that  $\Omega(z)$ (\ref{N_to_0}) provides the solution for the
initial recurrent system (\ref{System_RR}) and is analytic in a circle around
$z=0$. Therefore, inside this circle the solution is unique as its Taylor expansion coincides with that
generated by the initial recurrence system (2.12).
}

These simple solutions show that the bi-quartic theory
(\ref{LagrangianBissextile})
exhibits complicated structure in the leading logarithmic approximation.
It seems that, depending on the values of parameter $N$, this theory has qualitatively different phases. Further examples of an even more complicated analytical structure of the
LL-amplitude
are provided by the solutions in terms of elliptic (meromorphic, doubly periodic) functions.

\subsection{Solutions in terms of elliptic functions}
\label{EllipticFsol}

\begin{figure}[h!]
\centering
\includegraphics[height=9.0cm]{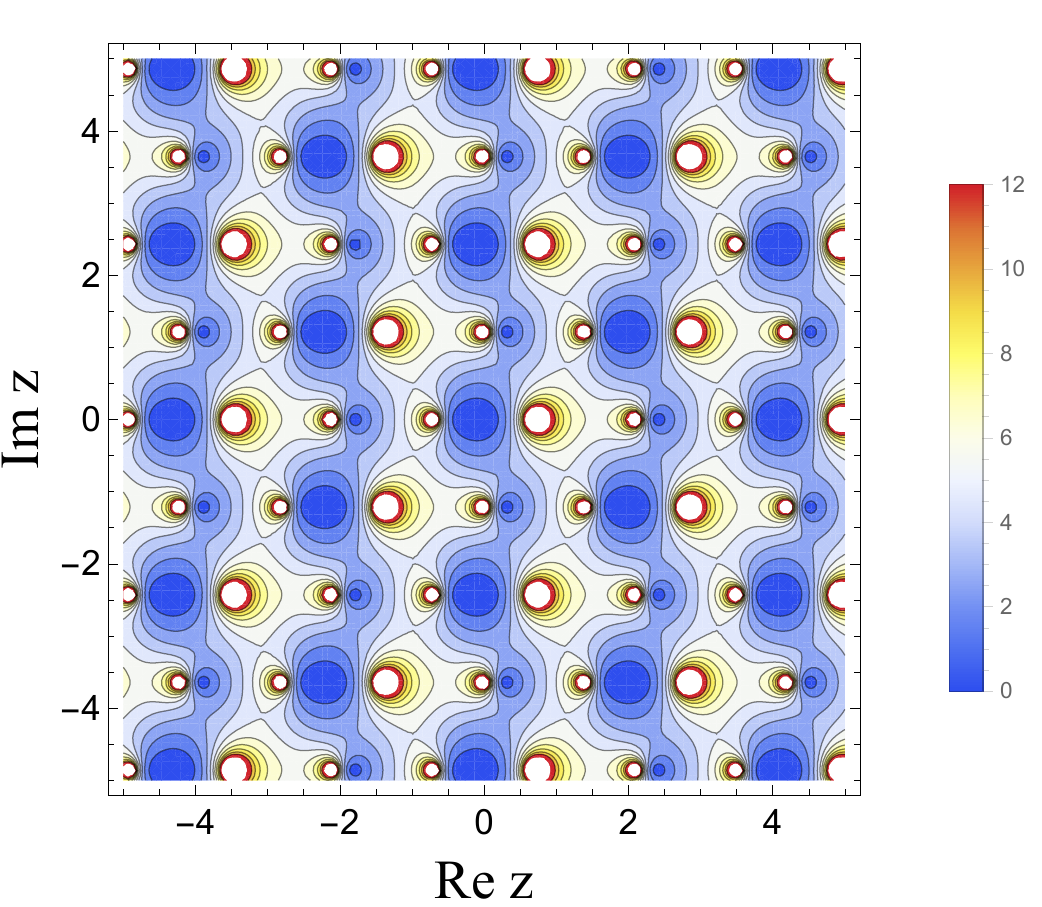}	
\caption{\label{Fig-1:processes}
	The contour plot of $|\Omega(z)|/(N-1)$ for the bi-quartic QFT (\ref{LagrangianBissextile}) in the $N\to 1$ limit.}
\end{figure}

Probably the most interesting case is the theory
(\ref{LagrangianBissextile})
in the limit of
$N\to 1$.
In terms of the effective mechanical system
(\ref{eq:mech_system})
it corresponds to  $\gamma=3$ and $m=\frac{1}{2}$.
The corresponding differential equation can be solved
in terms of the Weierstra{\ss} elliptic function.
The solution has the form:
\be
q(t)=\frac{3 \wp\left(\sqrt 3 t;{0,-\frac{4}{27}}\right)-2}{3 \wp\left(\sqrt 3 t;{0,-\frac{4}{27}}\right)+1}.
\ee
Here the Weierstra{\ss} elliptic function
$\wp(z;\, {\rm g_2,g_3})$
with the invariants
${\rm g}_2, {\rm g}_3$
satisfies the master differential equation%
\footnote{This definition is consistent with that employed in the Wolfram Mathematica$^\circledR$.}
\cite{VIS}:
\be
[\wp']^2=4 \wp^3-{\rm g}_2 \wp-{\rm g}_3, \quad \wp(z\to 0; \, {\rm g}_2,
{\rm g}_3)=\frac{1}{z^2}.
\ee
With the help of eq.~(\ref{eq:omS}) we obtain the LL-amplitude
$\Omega(z)$
in the limit
$N\to 1$:
\be
\label{eq:Bacher}
\frac{\Omega(z)}{N-1}= \frac{1}{2}\ \frac{6 \wp\left(\sqrt 3 z;{0,-\frac{4}{27}}\right)-\sqrt 3 \wp'\left(\sqrt 3 z;{0,-\frac{4}{27}}\right) +2 }{\wp\left(\sqrt 3 z;{0,-\frac{4}{27}}\right)^2-\frac 13 \wp\left(\sqrt 3 z;{0,-\frac{4}{27}}\right)-\frac 29}.
\ee
On Fig.~\ref{Fig-1:processes} we show the plot of $|\Omega(z)|/(N-1)$
as a function of the complex variable $z$.
One can see on this figure the periodically located poles (white areas) and zeros (deep blue areas) on the complex $z$-plane.
The detailed analysis of the solution
(\ref{eq:Bacher})
and its relation to pseudo-factorials, Dixon's elliptic functions, continued fraction, and to combinatorial analysis can be found in Refs.~\cite{Baher,Flajolet} (see also \cite{QRI}).

There are two additional cases in which
$\Omega(z)$
can be expressed in terms of elliptic functions.
Below we just list them:
\bi
\item \underline{$N=\frac{2}{3}$:}
\be
\Omega(z)=\frac{4 \wp\left(\frac 43 z;-1,0\right)-\wp'\left(\frac 43 z;-1,0\right)+2}{-3\wp\left(\frac 43 z;-1,0\right)^2+\frac34}.
\label{Elliptic_N=2/3}
\ee
This solution is a doubly periodic meromorphic function of the variable $z$.

\item \underline{$N=\frac{2}{5}$:}
\be
\Omega(z)&=&\frac{450 \wp
   '\left(z;0,\frac{128^2}{125^2}\right)}{625 \wp
   \left(z;0,\frac{128^2}{125^2}\right
   )^2+400 \wp
   \left(z;0,\frac{128^2}{125^2}\right
   )-512}\nn \\
   &-&\frac{18}{5}
   \left(\frac{1}{\sqrt{1-\frac{48}{25
   \wp
   \left(z;0,\frac{128^2}{125^2}\right
   )+32}}}-1\right).
   \label{Elliptic_N=2/5}
\ee
This solution is also doubly periodic in the complex $z$-plane;
it possesses an  infinite number of poles and branch 
points.
\ei

\subsection{Qualitative analysis of general case}
\label{Qual_analys}

Solutions for more general values of
$N$ can be obtained in terms of hyper-elliptic functions
(see, {\it e.g.} ref.~\cite{Kunz}).
Here, using the effective mechanical system
(\ref{eq:mech_system}),
we make several conclusions concerning the nature of the nearest singularity of
$\Omega(z)$
on the real and imaginary axes in the complex-$z$ plane.

For
\underline{
$N>0$
and
$N<-2$}
(corresponding to positive $\gamma$ in eq.~(\ref{eq:mech_system}))
the mechanical system describes the motion  of a ``particle'' to the left from
the initial position at the point $q=1$. In a finite time the ``particle'' reaches
the point $q=0$ with a finite ``velocity''. This point corresponds to a singularity of
the LL-amplitude $\Omega(z)$ (see eq.~(\ref{eq:omS})).
Moreover, since at this point the ``velocity'' is finite
the corresponding singularity is a pole.
Thus, the position of the nearest pole singularity on the real axis for $N>0$  and $N<-2$ can be easily computed with the result:
\be
z_{\rm pole}=\frac{\sqrt\pi \Gamma\left(1+\frac{N}{N+2}\right)}{2 N \Gamma\left(\frac12+\frac{N}{N+2}\right)}.
\label{Zp}
\ee

For the case of \underline{$-2<N<0$}
(corresponding to negative $\gamma$ in eq.~(\ref{eq:mech_system}))
the motion of the effective ``particle'' is to the right from
the initial point $q=1$. It never reaches the point $q=0$, {\it i.e.} the LL-amplitude
$\Omega(z)$ has no singularities on the real axis.

In the case $N=\frac{2}{2p+1}$ ($p\in \mathbb{N}$)
the effective ``particle''  oscillates in the potential
$q^{2 (p+1)}$.  Therefore, it periodically crosses the
point $q=0$ with a finite ``velocity''.
This point corresponds to a pole of
$\Omega(z)$.
The position of the pole nearest to the origin is given
by $z_{\rm pole}$
(\ref{Zp}).
On the real $z$ axis the LL-amplitude, therefore, possesses an infinite number of equidistant poles separated by
\be
\Delta z=\frac{(2 p+1)\sqrt\pi\ \Gamma\left(1+\frac{1}{2(p+1)}\right)}{2\ \Gamma\left(\frac12+\frac{1}{2(p+1)}\right)}, \quad p\in \mathbb{N}.
\ee
The singularities of the LL-amplitude
$\Omega(z)$
on the imaginary axis of $z$ can be obtained by considering the motion for effective mechanical
system (\ref{eq:mech_system}) in the imaginary time
$\tau=-i t$.
From these consideration one can conclude that
\bi
\item
for $-2<N<0$ the LL amplitude has the nearest branch 
point\footnote{The singularity is the branch 
point as the corresponding
``velocity" is infinite at $q=0$.} and the pole at
\be
z_{\rm branch}=\pm i \frac{\sqrt{\pi } \Gamma
   \left(\frac{2}{N+2}-\frac{1}{2}\right)}{2 N\ \Gamma
   \left(-\frac{N}{N+2}\right)}.
\ee
 \item for $|N|>2$ the LL amplitude $\Omega(z)$ has no singularities on the
 imaginary $z$ axis.
\ei

\section{Discussion and outlook}

We presented a general method for exact summation of  leading logarithms in 2D non-renormalizable EFTs.
The method was applied to the non-renormalizable EFT (\ref{LagrangianBissextile}).
This theory is the two-dimensional sibling of
the  familiar four-dimensional $\sigma$-models (like chiral EFTs).  For the first time, the all-order summation of the leading infra-red logarithms (chiral logs)
was performed in the closed analytical form.

Depending on the dimension of the internal symmetry space $N$, we obtained a number of exact solutions
for the LL scattering amplitudes with rich variety of analytical properties --in most of the cases the solutions are meromorphic
functions with infinite number of the Landau poles.
These cases provide the realization for the quasi-renormalizable QFTs discussed recently in ref.~\cite{QRI}. Generally the bi-quartic theory (\ref{LagrangianBissextile})
considered here possesses   many very different phases in the LL-approximation depending on the rank of the symmetry group.

Probably, one of the most interesting solutions corresponds to the EFT (\ref{LagrangianBissextile}) with  $N=2$.
In this case the LL amplitude:
\be
\Omega^{N=2}(z) =\frac{2}{\cos(4z)}+\tan(4 z)-2,
\ee
has an infinite set of poles equidistantly distributed on the real axis of $z$ . We can speculate that this feature reminds the properties of the Veneziano amplitude in the string theory, and there might be
deep relations between the properties of the LL-approximation and the mass spectrum in full theory.

Another remarkable case is the EFT (\ref{LagrangianBissextile}) in the limit of $N\to1$.
The corresponding LL-amplitudes turn to be doubly periodic meromorphic functions (elliptic functions). This case has deep relations to the theory of continued fractions
and combinatorial analysis, see \cite{Baher,Flajolet}. It would be extremely interesting to find
a real physical system that corresponds to the QFT (\ref{LagrangianBissextile}) in this limit.
We considered here only a particular bi-quartic QFT (\ref{LagrangianBissextile}), but the class of such theories is very wide, and the general method presented
in this paper can be applied to work out exact solutions of such theories in the LL-approximation.

\acknowledgments
We are grateful to D.~Kazakov, N.~Kivel, A.~Smirnov, and A.~Vladimirov for inspiring discussions.
MVP is grateful to K.~Efetov for interesting discussion on the relevance
of the bi-quartic theories for the description of the insulator-conductor phase transitions.
The work  is supported in parts by the BMBF (grant 05P2018) and CRC110 (DFG).

\appendix
\section{ Derivation of recurrence equations for LL-coefficients in 2D}
\label{AppA}

In this Appendix, following the method of ref.~\cite{Koschinski:2010mr}, relying
on the fundamental QFT requirements of unitarity, analyticity, and crossing symmetry, we derive the recurrence relation for leading IR logarithms in a general
$\Phi^4$-type theory in two dimensions defined by the action:
\be
S=\int d^2 x\ \left(\frac12 \partial_\mu\Phi^a \partial^\mu\Phi^a -V(\Phi,\partial\Phi)\right).
\ee
Here the expansion of the interaction
$V(\Phi,\partial\Phi)$
starts with four fields $\Phi$, the corresponding $\Phi^4$-part contains $2\kappa$ derivatives.
The theories in which the number of derivatives is a multiple of four we call as {\it bi-quartic}.
The index $a$ (it can be a multi-index as well) corresponds to a possible internal symmetry of the theory. The bi-quartic theory
(\ref{LagrangianBissextile})
belongs to this class of theories with $\kappa =2$ and the index $a$ corresponding to the vector representation of the  ${\rm O}(N)$ isotopic group.

\begin{figure}[th]
\centerline{\includegraphics[scale=0.50]{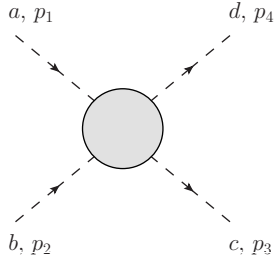}}
\vspace*{8pt}
\caption{Assignment of momenta and group indices for the scattering amplitude (\ref{2to2scattering}).
\label{Fig:2to2}}
\end{figure}

We consider the $2 \to 2$ scattering reaction (see Fig.~\ref{Fig:2to2} for assignment of momenta and group indices)
\be
 \Phi_a(p_1)+\Phi_b(p_2)\to \Phi_c(p_3)+\Phi_d(p_4)
 \label{2to2scattering}
 \ee
in two space-time dimensions. We employ the following parametrization for the involved particle momenta:
\be
p= (E({\bf p}),{\bf p}), \ \ {\rm with} \ \ E({\bf p})=\sqrt{{\bf p}^2+m^2}.
\ee
Throughout the derivation we keep the mass of particles non-zero and put them to zero at the very last step.
We adopt the standard normalization of one-particle states:
\be
\langle \Phi_a(p')| \Phi_b (p) \rangle= 2 E  (2 \pi) \delta( {\bf p} - {\bf p'}) \delta_{ab}.
\ee

The  scattering matrix element for the
reaction (\ref{2to2scattering})
reads
\be
&&
\langle \Phi_c(p_3) \Phi_d(p_4) |S|  \Phi_a (p_1) \Phi_b (p_2) \rangle
\nonumber \\ && = \mathbb{I}+ i (2 \pi)^2 \delta({\bf p_3}+{\bf p_4}-{\bf p_1}-{\bf p_2})
\delta(E_3+E_4-E_1-E_2) \sum_{I} P^I_{abcd} \, {\cal M}^I(s,t,u),
\label{Smatrix2to2}
\ee
where
\be
\mathbb{I}=(2 \pi)^2 2E_1 2 E_2  \delta({\bf p_1}-{\bf p_4}) \delta({\bf p_2}-{\bf p_3}) \ \delta_{ad} \delta_{bc},
\ee
and the sum $\sum_{I} $ runs over all irreducible representation of the symmetry group in the $s$-channel.
In the 2D case the description of the reaction
(\ref{2to2scattering})
in terms of the Mandelstam variables $s$, $t$, $u$ is somewhat redundant. Due to the vanishing
phase space there are only two physical possibilities for
scattering (we adopt the center-of-mass frame):
\begin{enumerate}
\item Forward scattering
$
t=0; \ \ \ u=4m^2-s.
$
\item Backward scattering
$
u=0; \ \ \ t=4m^2-s.
$
\end{enumerate}

We employ the following expression for the overall momentum conservation
delta function in (\ref{Smatrix2to2}) \cite{Delphenich:1998ht}:
\be
&&
\delta({\bf p_3}+{\bf p_4}-{\bf p_1}-{\bf p_2})
\delta(E_3+E_4-E_1-E_2) \nonumber \\ &&
=\left| \frac{\partial E_1({\bf p_1})}{\partial{\bf p_1} } -
\frac{\partial E_2({\bf p_2})}{\partial{\bf p_2} } \right|^{-1}
\big[
\delta( {\bf p_1}-{\bf p_4}) \delta( {\bf p_2}-{\bf p_3})+
\delta( {\bf p_1}-{\bf p_3}) \delta( {\bf p_2}-{\bf p_4})
\big]\nonumber \\ &&
= \frac{s}{2 \sqrt{s(s-4m^2)}} \big[
\delta( {\bf p_1}-{\bf p_4}) \delta( {\bf p_2}-{\bf p_3})+
\delta( {\bf p_1}-{\bf p_3}) \delta( {\bf p_2}-{\bf p_4})
\big].
\ee
Therefore, we can rewrite the matrix element (\ref{Smatrix2to2})  as
the sum of ``transmission'' ($T$) and ``reflection'' ($R$) pieces:
\be
&&
\langle \Phi_c(p_3) \Phi_d(p_4) |S|  \Phi_a (p_1) \Phi_b (p_2) \rangle
 = \mathbb{I}   + i (2 \pi)^2  \frac{s}{2 \sqrt{s(s-4m^2)}} \nonumber \\ && \times \Big[
\delta( {\bf p_1}-{\bf p_4}) \delta( {\bf p_2}-{\bf p_3})
\sum_{I} P^I_{abcd} \, {\cal M}^{I, \, T}(s)
+
\delta( {\bf p_1}-{\bf p_3}) \delta( {\bf p_2}-{\bf p_4})
\sum_{I} P^I_{abcd} \, {\cal M}^{I, \, R}(s) \Big].
\nonumber \\ &&
\label{Smatrix2to2_TR}
\ee
Note that the transmission and reflection amplitudes depend now only on one Mandelstam variable $s$.

We need to work out the $2$-particle unitarity relation for
$2 \to 2$
scattering in $1+1$ dimensions:
\be
&&
2\ {\rm Im} \langle    p_3    p_4| {\cal M}|    p_1    p_2 \rangle= \frac{1}{2 ! }  \int \frac{d {\bf p_1'}}{(2\pi) 2 E_1'}
\frac{d {\bf p_2'}}{(2\pi)2 E_2'}
\langle    p_1    p_2| {\cal M}^\dag |   p_1'   p_2'\rangle \,
\langle    p_3    p_4| {\cal M}  |   p_1'    p_2'\rangle \,
\ee
Therefore,
\be
&&
{\rm Im} {\cal M}^T(s) \Big|_{s>0}= \frac{1}{8 s}\ \frac{s}{ \sqrt{s(s-4m^2)}} \left( |{\cal M}^T(s)|^2+|{\cal M}^R(s)|^2 \right); \nonumber \\ &&
{\rm Im} {\cal M}^R(s) \Big|_{s>0} = \frac{1}{8 s}\ \frac{s}{ \sqrt{s(s-4m^2)}} \left( {{\cal M}^{T}}^*(s){\cal M}^R(s)+
{{\cal M}^R}^*(s){\cal M}^T(s) \right).
\ee
Let us consider the most general form of the LL-approximation for
the transmission and reflection amplitudes ${\cal M}^{I, \, T}$, ${\cal M}^{I, \, R}$ introduced in
eq.~(\ref{Smatrix2to2_TR})
\be
&&
{\cal M}^{I, \,T}\Big|_{\rm LL}(s)= 4 \pi s \sum_{n=1}^\infty \hat{S}^n \sum_{i=0}^{n-1} \alpha_{n,i}^{I, \, T}
\ln^i \left( \frac{\mu^2}{s} \right) \ln^{n-i-1} \left( \frac{\mu^2}{-s} \right)+
\cal{O}({\rm NLL}); \nonumber \\ &&
{\cal M}^{I, \,R}\Big|_{\rm LL}(s)=  4 \pi s \sum_{n=1}^\infty \hat{S}^n \sum_{i=0}^{n-1} \alpha_{n,i}^{I, \, R}
\ln^i \left( \frac{\mu^2}{s} \right) \ln^{n-i-1} \left( \frac{\mu^2}{-s} \right)+
\cal{O}({\rm NLL}).
\label{TR_amplitudes_LLogs}
\ee
Here $\hat{S}$ stands for the dimensionless expansion parameter
\be
\hat{S}= \frac{s^{\kappa -1}}{4\pi F^2},
\ee
where $2\kappa$ is the number of derivatives in the interaction part of the Lagrangian.
$1/F^2$
is the corresponding coupling constant with the dimension
$[F]=\kappa-1$.
Note that $\kappa=1$ corresponds to a renormalizable theory in 2D.

In analogy with the general reasoning of
ref.~\cite{Koschinski:2010mr},
we would like to work out the recurrence relations for the LL-coefficients of
amplitudes
(\ref{TR_amplitudes_LLogs}).
Obviously, one has
\be
\omega_{n}^{I, \,T}= \sum_{i=0}^{n-1}  \alpha_{n,i}^{I, \, T}; \ \ \
\omega_{n}^{I, \,R}= \sum_{i=0}^{n-1}  \alpha_{n,i}^{I, \, R};
\ee
Unitarity relation for $s>0$ (right cut) gives
\be
&&
\sum_{i=0}^{n-1} \alpha_i^{I,\,T} (n-i-1)= \frac 12 \sum_{k=1}^{n-1}  \left( \omega_k^{I,\,T} \omega_{n-k}^{I,\,T}+
\omega_k^{I,\,R} \omega_{n-k}^{I,\,R} \right); \nonumber \\ &&
\sum_{i=0}^{n-1} \alpha_i^{I,\,R} (n-i-1)= \frac 12 \sum_{k=1}^{n-1}  \left( \omega_k^{I,\,T} \omega_{n-k}^{I,\,R}+
\omega_k^{I,\,R} \omega_{n-k}^{I,\,T} \right),
\label{From_right_cut}
\ee
where, we employ that, according to our conventions,
\be
&&
{\rm Im} \left( \ln \frac{\mu^2}{-(s+i \varepsilon)} \right)^{n-i-1}= {\rm Im} \left(i \pi+ \ln \frac{\mu^2}{s} \right)^{n-i-1}
\nonumber \\ &&
=\pi (n-i-1) \left( \ln \frac{\mu^2}{s} \right)^{n-i-2}+ {\cal O}({\rm NLL \; terms}).
\ee

The dispersion relations for the transmission and reflection amplitudes have the following form:
\be
&&
{\cal M}^I(s, t=0, u=4m^2-s) \equiv {\cal M}^{I,\,T}(s )
\nonumber \\ &&
= \frac{1}{\pi} \int_{4m^2}^\infty
ds' \sum_J \left(\frac{\delta^{IJ }}{s'-s}+ \frac{C_{su}^{IJ }}{s'-u} \right) {\rm disc}   {\cal M}^{I,\,T}(s' );
\label{DR_transmission}
\ee
and
\be
&&
{\cal M}^I(s, t=4m^2-s, u=0) \equiv {\cal M}^{I,\,R}(s )
\nonumber \\ &&
= \frac{1}{\pi} \int_{4m^2}^\infty
ds' \sum_J \left(\frac{\delta^{IJ }}{s'-s}+ \frac{C_{st}^{IJ }}{s'-t} \right) {\rm disc}   {\cal M}^{J ,\,T}(s' ).
\label{DR_reflection}
\ee
Here where the crossing matrices $C_{xy}^{IJ }$ were defined for a general symmetry group%
\footnote{For the case of the $O(N)$group they are given by eq.~(\ref{CrossM}).}
in ref.~\cite{Koschinski:2010mr}.

Now we employ the equations
(\ref{DR_transmission}), (\ref{DR_reflection}) to work out the
unitarity relation on l.h.s. cut:
\be
{\rm Im} {\cal M}^{I,\,T}(s) \Big|_{s<0}&=& \sum_J C_{su}^{IJ }  \frac{1}{8 s}\ \frac{s}{ \sqrt{-s(-s-4m^2)}} \left( |{\cal M}^{J ,\,T}(-s)|^2+|{\cal M}^{J ,\,R}(-s)|^2 \right)
\\
{\rm Im} {\cal M}^{I,\,R}(s)\ \Big|_{s<0}
&=& \sum_J C_{st}^{IJ } \frac{1}{8 s} \frac{s}{ \sqrt{-s(-s-4m^2)}} \left( {{\cal M}^{J ,\, T}}^*(-s){\cal M}^{J , \,R}(-s)+
{{\cal M}^{J ,\,R}}^*(-s){\cal M}^{J ,\,T}(-s) \right).\nn
\ee
Therefore, we get
\be
&&
\sum_{i=0}^{n-1} \alpha_i^{I,\,T} (i)= -\frac{(-1)^{(\kappa-1)n}}{2}  \sum_{k=1}^{n-1} \sum_J C_{su}^{IJ } \left( \omega_k^{J ,\,T} \omega_{n-k}^{J ,\,T}+
\omega_k^{J ,\,R} \omega_{n-k}^{I,\,R} \right); \nonumber \\ &&
\sum_{i=0}^{n-1} \alpha_i^{I,\,R} (i)= -\frac{(-1)^{(\kappa-1)n}}{2}  \sum_{k=1}^{n-1}  \sum_J C_{st}^{IJ } \left( \omega_k^{J ,\,T} \omega_{n-k}^{J ,\,R}+
\omega_k^{J ,\,R} \omega_{n-k}^{J ,\,T} \right).
\label{From_left_cut}
\ee

Combining
(\ref{From_right_cut})
and
(\ref{From_left_cut})
we establish the recurrence relations for the LL-coefficients  in 2D:
\be
&&
\omega_{n}^{I, \,T}= \frac{1}{2(n-1)}
\sum_{k=1}^{n-1} \sum_J
\Big(
\delta^{IJ }- (-1)^{(\kappa-1)n} C_{su}^{IJ }
\Big)
\left( \omega_k^{J ,\,T} \omega_{n-k}^{J ,\,T}+
\omega_k^{J ,\,R} \omega_{n-k}^{J ,\,R} \right);
\nonumber \\ &&
\omega_{n}^{I, \,R}= \frac{1}{2(n-1)}
\sum_{k=1}^{n-1} \sum_J
\Big(
\delta^{IJ }- (-1)^{(\kappa-1)n} C_{st}^{IJ }
\Big)
\left( \omega_k^{J ,\,T} \omega_{n-k}^{J ,\,R}+
\omega_k^{J ,\,R} \omega_{n-k}^{J ,\,T} \right).
\label{MainRecRel}
\ee
Note that for even
$\kappa$
(once the number of derivatives in the interaction is a multiple of $4$)
the recursive equation has the form which is different from the usual
RG-equation type due to the presence of $(-1)^n$ factor in the equation.
Therefore, in this case the recurrence equations can have solutions with  rather complicated analytical properties, see detailed discussion in \cite{QRI}.
We call such special type of QFTs as {\it bi-quartic theories}. These theories turn
to be natural candidates for quasi-renormalizable QFTs introduced in \cite{QRI}.

\end{document}